\documentstyle[epsf,epsfig]{article}
\begin{document}
\normalsize
\large
\begin{center}
  \bf
  Test of Time Reversal Invariance  \\
  in Polarized Proton-Deuteron Scattering\footnote{CONTRIBUTION TO THE
    FIFTH INTERNATIONAL WEIN SYMPOSIUM: A CONFERENCE ON PHYSICS BEYOND
    THE STANDARD MODEL}

\end{center}
\normalsize
\begin{center}
F. Hinterberger for the TRI collaboration\footnote{J. Bisplinghoff$^{1}$, 
 J. Ernst$^{1}$,
 P.D. Eversheim$^{1}$,
 F. Hinterberger$^{1}$, 
 R. Jahn$^{1}$, 
 H.E. Conzett$^{2}$, 
 M. Beyer$^{3}$,
 H. Paetz gen. Schieck$^{4}$,
 W. Kretschmer$^{5}$,  Spokesman P.D. Eversheim$^{1}$\\
\noindent
$^1$Inst. f. Strahlen- und Kernphysik, Univ. Bonn, Germany\\
$^2$Lawrence Berkeley Laboratory, Berkeley, USA,\\
$^3$FB Physik, Univ. Rostock, Germany,\\
$^4$Inst. f. Kernphysik, Univ. K\"oln, Germany,\\
$^5$Physik. Inst., Univ. Erlangen, Germany} at COSY \\
Inst. f. Strahlen- und Kernphysik, Univ. Bonn, Germany\\
\end{center}

\begin{abstract}
  A novel test of time-reversal invariance in proton-deuteron
  scattering is planned as an internal target transmission experiment
  at the cooler synchrotron COSY.  The P-even, T-odd observable is the
  polarization correlation $A_{y,xz}$ of the total cross section
  measured using a polarized internal proton beam (polarization $p_y$)
  and an internal polarized deuterium target (tensor polarization
  $p_{xz}$).  Measuring this observable is a true null test of time
  reversal invariance and therefore allows to reach a high accuracy.
  Sufficient luminosity can be obtained using a window-less storage
  cell placed on the axis of the proton beam. Tensor polarized atoms
  are produced in an atomic beam source based on Stern-Gerlach
  separation in permanent sextupole magnets and adiabatic high
  frequency transitions.  The total cross section correlation is
  measured by monitoring the beam transmission in the COSY storage
  ring mode of operation.  The proton beam momentum will be in the
  range 2-3 GeV/c. This momentum is ideally suited to test possible
  short range contributions, i.e. natural parity charged $\rho$-type
  and unnatural parity $a_1$-type meson exchange contributions.  The
  feasibility of the experiment, systematic errors and the expected
  accuracy are discussed.
\end{abstract}
\section{Introduction}

The CPT-theorem \cite{fhint:lud54} connects the symmetry operation of
time reversal T with the particle-antiparticle and space inversion CP.
The existence of CP-violation is well established through neutral kaon
decays \cite{fhint:chr64}.  Assuming CPT-conservation CP-violation
implies also T-violation.  In nuclear reactions detailed balance,
polarization-analyzing power (P-A) tests and charge symmetry breaking
(CSB) tests in neutron-proton scattering have been performed yielding
upper limits for T-odd interactions. However because of the
requirement to compare a reaction observable to an observable in the
inverse reaction the experimental accuracy is limited to a level of
10$^{-2}$ -10$^{-3}$. The accuracy can be increased by several orders
of magnitude if a null test is performed, i.e. a single observable is
measured which must be zero by time reversal invariance (TRI).  As
shown by Conzett \cite{fhint:con93} the correlation $A_{y,xz}$ of the
spin 1/2 polarization $p_y$ and the spin 1 tensor polarization
$p_{xz}$ in the proton-deuteron total cross section is such an
observable.  The aim of the present experiment is to measure this
spin-dependent total cross section with an accuracy of $10^{-6}$.

In discussing tests of time reversal invariance (TRI)
one distinguishes parity violating (P-odd) from
parity conserving (P-even) time reversal noninvariant (T-odd)
interactions. 
The most precise constraints on P-odd/T-odd interactions come
from upper limits of the electric dipole moment of the
neutron and the atoms $^{129}$Xe  and
$^{199}$Hg providing a limit on a P-odd/T-odd pion-nucleon
coupling constant which is less than 10$^{-4}$ times the 
weak interaction strength.
Constraints  on P-odd/T-odd and P-even/T-odd
interactions are not independent since 
weak corrections to P-even/T-odd interactions can
generate P-odd/T-odd observables. 
Therefore the electric dipole measurements provide also upper limits
on the P-even/T-odd coupling strengths.
In a recent analysis a limit on the P-even/T-odd
$\rho$NN coupling strength,
$g^{\rm T}_{\rho{\rm NN}}/g_{\rho{\rm NN}}\le 1\cdot 10^{-3}$,
was deduced \cite{fhint:hax94}.

Direct experimental limits on P-even/T-odd interactions
are much less stringent. Detailed balance tests  
of the reactions $^{24}$Mg($\alpha$,p)$^{27}$Al and its inverse
\cite{fhint:bla83}
yield an accuracy of $\Delta =5.1\cdot10^{-3}$ (80\% CL) and
$\alpha_{\rm T} \le 2\cdot 10^{-3}$ \cite{fhint:hax94}
where $\alpha_{\rm T}$ measures the ratio of T-odd to T-even
nuclear matrix elements. A recent P-even/T-odd TRI test
of the forward scattering amplitude
using polarized neutrons and nuclear spin aligned Holmium
yielded $\alpha_{\rm T} \le 7\cdot 10^{-4}$  \cite{fhint:huf96}
corresponding to a bound of the P-even/T-odd $\rho$NN coupling constant
$g^{\rm T}_{\rho{\rm NN}}/g_{\rho{\rm NN}}\le 6\cdot 10^{-2}$.
Simonius \cite{fhint:sim97} analyzed
charge symmetry breaking (CSB) tests
in neutron-proton scattering and deduced 
$g^{\rm T}_{\rho{\rm NN}}/g_{\rho{\rm NN}}\le 6.7\cdot 10^{-3}$.

\section{Effective T-odd NN Interactions}
In order to analyze TRI tests at low and intermediate energies
effective meson exchange potentials may be used \cite{fhint:bey93}.
For P-odd/T-odd interactions the longest range potential may
be parametrized as a $\pi$-exchange potential,
\begin{equation}
V^{\rm PT}_{\pi}=\phi^{\rm PT}_{\pi}
\frac{g_{\pi NN}^2}{2m_p(\vec{q}^2+m_{\pi}^2)}
(\vec{\sigma}_1-\vec{\sigma}_2)\cdot\vec{q}\; \;
 (\vec{\tau}_1\cdot \vec{\tau}_2)
\end{equation}
with $\vec{q}=\vec{p}_f-\vec{p}_i$ and $\phi^{\rm PT}_{\pi}$ the
P-odd/T-odd coupling strength.

As shown by Simonius\cite{fhint:sim75} P-even/T-odd nucleon-nucleon
interaction can only occur for J$\neq$0 single meson exchange.  Thus,
there is no long-range pion exchange possible.  For natural parity
$\pi =(-1)^{\rm J}$ exchanges (J$^\pi$=1$^-$, 2$^+$,...)  the meson
must be charged (e.g. $\rho^{\pm}$ exchange) and can only contribute
to the np interaction in singlet-triplet transitions $^1{\rm P}_1
\leftrightarrow {^3{\rm P}_1}$, $^1{\rm D}_2 \leftrightarrow {^3{\rm
    D}_2}$, etc.  T-violation is due to the isospin operator
$[\vec{\tau}_1\times \vec{\tau}_2]_3={\rm i} (\tau_1^+ \tau_2^- -
\tau_1^- \tau_2^+)$.  For unnatural parity $\pi =(-1)^{{\rm J}-1}$
exchanges (J$^\pi$=1$^+$, 2$^-$,...)  like the a$_1$(J$^{\pi}$=1$^+$)
exchange there is no charge restriction. They may contribute to the np
as well as the nn and pp interaction.  T-violation must be in the
spin-space operator. For nn and pp the lowest partial wave transition
is $^3{\rm P}_2 \leftrightarrow {^3{\rm F}_2}$.  The P-even/T-odd
potentials may be written
\begin{eqnarray}
V^{\rm T}_{\rho} & = & i\phi^{\rm T}_{\rho}
\frac{\kappa_{\rho}g_{\rho NN}^2}{8m_p^2(\vec{q}^2+m_{\rho}^2)}
(\vec{\sigma}_1-\vec{\sigma}_2)\cdot\vec{q}\times \vec{p}\; \; 
[\vec{\tau}_1\times \vec{\tau}_2]_3 \\
V^{\rm T}_{{\rm a}_1} & = & i\phi^{\rm T}_{{\rm a}_1}
\frac{g_{{{\rm a}_1} {\rm NN}}^2}{8m_p^2(\vec{q}^2+m_{{\rm a}_1}^2)}
(\vec{\sigma}_1\cdot \vec{p}\; \vec{\sigma}_2\cdot \vec{q}+
\vec{\sigma}_2\cdot \vec{p}\; \vec{\sigma}_1\cdot \vec{q}-
\vec{\sigma}_1\cdot \vec{\sigma}_2 \;  \vec{q}\cdot \vec{p}).
\end{eqnarray}
with $\vec{p}=(\vec{p}_f+\vec{p}_i)/2$ and $\phi^{\rm T}_{\rho}$ 
and $\phi^{\rm T}_{{\rm a}_1}$ the
P-even/T-odd coupling strengths.

The proton-deuteron system was analyzed by Beyer \cite{fhint:bey93} in
terms of effective P-even/T-odd nucleon-nucleon interactions.  He
showed that measuring $\sigma_{tot}$ with an experimental accuracy of
$1\cdot 10^{-6}$ yields tight bounds on the P-even/T-odd $\rho$NN and
${\rm a}_1$NN coupling constants, $\phi^{\rm T}_{\rho}=g^{\rm
  T}_{\rho{\rm NN}}/g_{\rho{\rm NN}}\le 1\cdot 10^{-3}$ and $\phi^{\rm
  T}_{{\rm a}_1}=g^{\rm T}_{{\rm a}_1{\rm NN}}/g_{{\rm a}_1{\rm
    NN}}\le 2\cdot 10^{-3}$.

\section{P-even/T-odd Observable}
The total cross section $\sigma_{\rm tot}$ involving polarized
particles is described by the generalized optical
theorem\cite{fhint:phi63}
\begin{equation}
\sigma_{\rm tot} = \frac{4\pi}{k} {\rm Im} \frac{{\rm Tr} 
\rho F(0)}{{\rm Tr} \rho}.
\end{equation}
Here, $\rho$ is the polarization density matrix of the
initial state, $k$ the cm-wave number and $F(0)$ the
matrix of the spin-dependent forward scattering amplitudes.
\begin{figure}[t]
\begin{center}
\psfig{figure=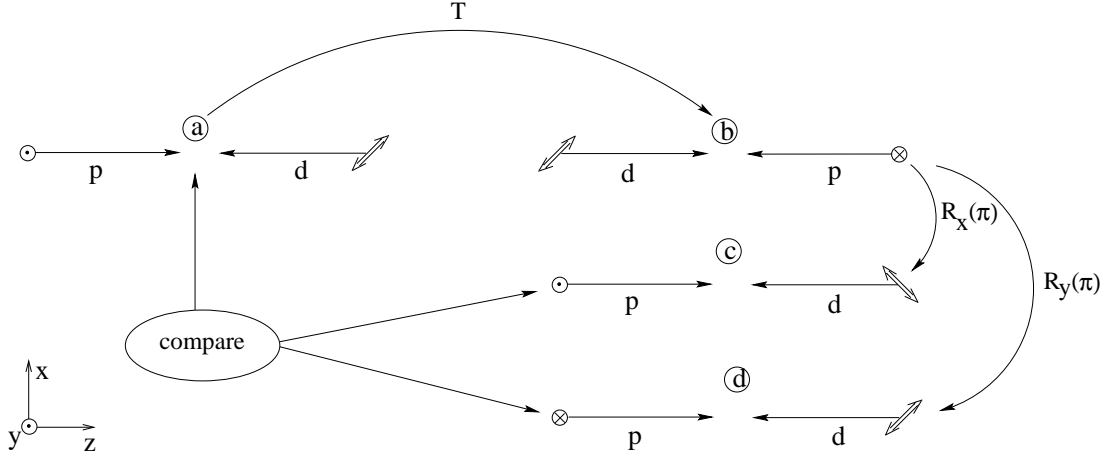,height=6cm}
\end{center}
%\vspace {3 cm}
\caption{Pictorial demonstration of T-odd polarization
correlation in pd forward scattering. System b is timereversed to
a. For a direct comparison system b is rotated through
180$^{\circ}$ about the x- and y-axis yielding system c and d,
respectively. The arrows denote cm momenta of the
incoming particles p and d. The
symbols $\bigodot$ and $\bigotimes$ denote positive and
negative spin polarization $p_y$ of the proton,
the symbol $\Longleftrightarrow$ tensor polarization
$p_{xz}$ of the deuteron.}
%\label{fig:edda}
\end{figure}

For a P-even/T-odd experiment one may then choose 
polarized protons (polarization $p_y$) and tensor polarized
deuterons (tensor polarization $p_{xz}$) (see fig.1), 
and the total cross section has a P-even/T-odd term
\begin{equation}
\sigma_{\rm tot} = \sigma_{\rm tot}^0 (1+  
A_{y,xz} p_y p_{xz}).
\end{equation}
Here, $\sigma_{\rm tot}^0$ is the total cross section for
unpolarized beam and unpolarized target and 
$A_{y,xz}$ is the P-even/T-odd polarization correlation.
The number $N$ of stored beam particles decreases exponentially
with time $t$,
\begin{equation}
N(t)=N_0\exp -[(\sigma_{\rm tot} + \sigma_{\rm loss})\rho f t]
=N_0\exp -\lambda t.
\end{equation}
Here, $\rho$ is the arial density of the target, $\sigma_{\rm loss}$
the loss cross-section taking beam losses outside the target into
account, $f$ the revolution frequency of the circulating beam, $t$ the
time and $\lambda$ is the corresponding decay rate of the beam.  The
decay rate depends on the signs of $p_y$ and $p_{xz}$.  By measuring
in a sequence the decay rates $\lambda^{++}$, $\lambda^{+-}$ and
$\lambda^{-+}$ and $\lambda^{--}$ (see fig. 2, the superscripts refer
to the signs of $p_y$ and $p_{xz}$, respectively) and assuming
$|p_y^+|=|p_y^-|$ and $|p_{xz}^+|=|p_{xz}^-|$ one finally obtains
\begin{equation}
A_{y,xz} =\frac{\sqrt{\lambda^{++} \lambda^{--}}
-\sqrt{\lambda^{+-} \lambda^{-+}}}
{\sqrt{\lambda^{++} \lambda^{--}}+\sqrt{\lambda^{+-} \lambda^{-+}}}
\frac{1}{|p_y p_{xz}|} \left(1+\frac{\sigma_{\rm loss}}
{\sigma_{\rm tot}^0}\right).
\label{e.4} 
\end{equation}
\begin{figure}[t]
\begin{center}
\psfig{figure=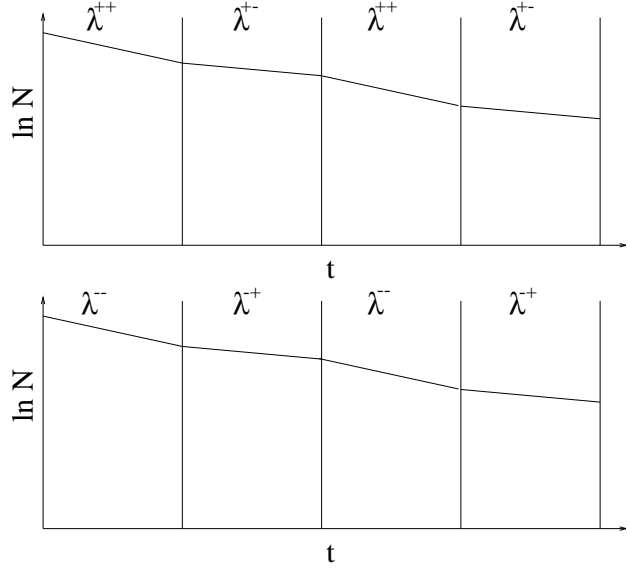,height=6cm}
\end{center}
%\vspace {3 cm}
\caption{Sequence of decay rate measurements.}
%\label{fig:edda}
\end{figure}

\section{Experiment}
The experiment is planned to be performed as an internal target
experiment in the cooler synchrotron COSY \cite{fhint:mai94}.  Thus, a
pure but in contrast to solid targets low density polarized atomic
beam target can be used.  The low target density is compensated by the
principle of recycling the beam with a frequency of about 1.5 10$^{6}$
s$^{-1}$. The experimental scheme is shown in fig. 3.  Basically the
experiment needs equipment that is already provided for the
EDDA-experiment\cite{fhint:bis91} at COSY, i.e.  the measurement of
$\vec{\rm p}$$\vec{\rm p}$ elastic scattering excitation functions.
Tensor polarized deuterium atoms are produced in an atomic beam source
based on Stern-Gerlach separation in permanent sextupole magnets and
adiabatic Abragam-Winter RF-transitions.  In order to increase the
luminosity the tensor polarized deuterium atoms are stored in a
windowless storage cell placed on the axis of the proton beam. The
tensor polarization is monitored by a spin filter located in the beam
dump of the polarized atomic beam target.
\begin{figure}[t]
\begin{center}
\psfig{figure=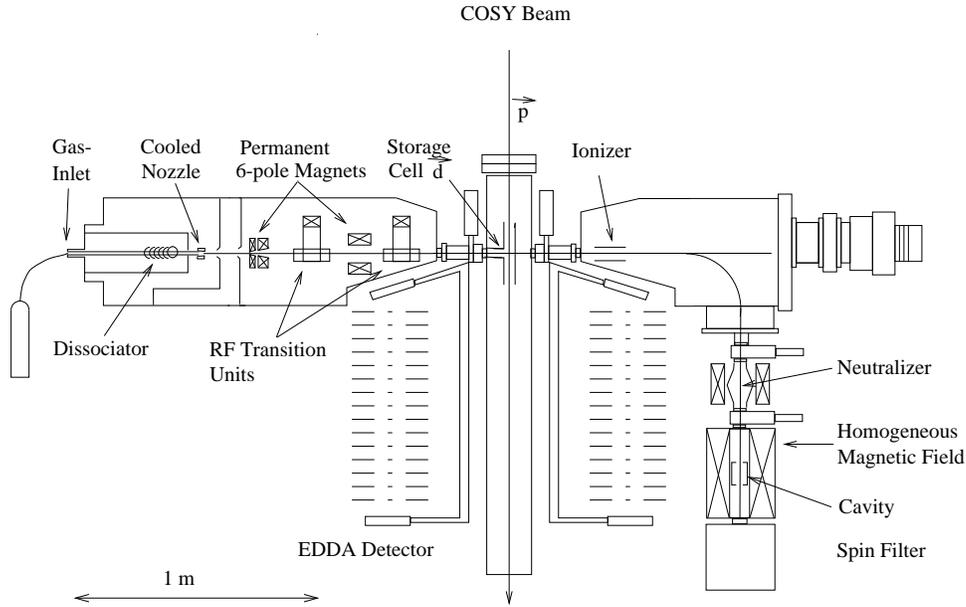,height=8cm}
\end{center}
%\vspace {3 cm}
\caption{Scheme of polarized atomic beam target and EDDA detector.}
%\label{fig:edda}
\end{figure}

The quantization (3) axis of the polarized atomic beam source is an
axis of cylindrical symmetry and the vector and tensor components
are $p_3$ and $p_{33}$. For
the measurements a pure $p_{33}=+1$ polarization state is
prepared. For a holding field  along the
direction $x=\pm z$ the resulting tensor components are:
$p_{xz}=\pm 3/4$, $p_{xx}=+1/4$, $p_{yy}=-1/2$ and $p_{zz}=+1/4$.
Thus, while flipping $p_{xz}$ the components $p_{xx}$,
$p_{yy}$ and $p_{zz}$ stay constant.

The statistical accuracy of the measurements depends on the luminosity.
With a target areal density of about 5 10$^{13}$ cm$^{-2}$,
10$^{11}$ polarized protons in the ring and a revolution frequency
of 1.5 10$^6$ s$^{-1}$ the luminosity will be
7.5 10$^{30}$ cm$^{-2}$s$^{-1}$. Thus, a statistical accuracy
of about 10$^{-6}$ can be reached in a 10 day run.

The signs of the beam and target polarizations $p_y$ and $p_{xz}$ are
chosen on a random basis.  A standard sequence of the experiment will
be: (i) The polarized proton beam is injected into the COSY ring and
accelerated to the appropriate energy.  (ii) The decay rate of the
beam is measured in the storage mode of operation by counting the
number of protons as a function of time.  (iii) The tensor
polarization of the target is flipped by an appropriate change of the
holding field.  (iv) The decay rate of the beam is now measured with a
flipped tensor polarization.  (v) The beam is decelerated and dumped.
Step (iii) and (iv) may be repeated several times.  The time period
for a single decay rate measurement depends on the statistical
accuracy. It will be in the order of 100 s.  So, in a 10 day run about
10$^4$ single measurements can be performed.

\section{Systematic errors}

Beam losses in the ring are harmless as long as they are smaller than
the losses due to the target, i.e.  $\sigma_{\rm loss} \leq
\sigma_{\rm tot}^0$.  They can be easily measured and corrected (see
eq. \ref{e.4}).  The holding field of the polarized deuterium target
causes a small distortion of the closed orbit. Fortunately a
correlation error can be avoided since the transverse component of the
holding field is not changed while flipping the tensor polarization.
Correlations between the phase space distribution of the stored beam
and the sign of the beam polarization are expected to be very small.
They are cancelled by flipping the tensor polarization of the target.

The most severe source of systematic errors are competing polarization
correlations.  Fortunately many polarization correlations, especially
$A_{y,xx}$, $A_{y,yy}$ and $A_{y,zz}$, vanish in forward scattering,
since they are odd with respect to a rotation about the $z$-axis.  In
addition the invariant spin axis of the COSY ring is in the
$y$-direction and the proton polarization vector $(0,p_y,0)$ is an
eigenvector of the ring.  Therefore, effects due to the polarization
components $p_x$ and $p_z$ are negligible.  Several correlations like
for instance $A_{y,x}$ violate parity conservation and are therefore
expected to be negligible (order 10$^{-7}$).  The most dangerous
polarization correlation is $A_{y,y}$ yielding a correlation between
beam and target vector polarization.  Therefore, a pure tensor
polarized deuterium target with negligible vector polarization is
prepared.  In addition a precise alignment of the target holding field
with respect to the $y$-axis is of great importance.  Summarizing, it
can be shown that all systematic errors can be kept below the
10$^{-6}$ level.

%\begin{references}

\end{document}